\documentclass[fleqn,twoside]{article}
\usepackage{espcrc2,cite,graphicx,amsmath,amsfonts,amssymb,mathrsfs}

\newcommand{\ie}{{\it i.e.}}

\newcommand{\eg}{{\it e.g.}}

\newcommand{\cf}{{\it cf.}}

\newcommand{\eq}{Eq.}
\newcommand{\eqs}{Eqs.}
\newcommand{\fig}{Fig.}
\newcommand{\figs}{Figs.}
\newcommand{\Ref}{Ref.}
\newcommand{\Refs}{Refs.}

\newcommand{\App}{App.}

\begin{document}

\title{
\vspace*{-3cm}
\begin{flushright}
{\small TUM-HEP-418/01}
\end{flushright}
{\bf Reconstruction of the Earth's matter density profile using a
single neutrino baseline}}

\author{{\large Tommy Ohlsson}\address[TUM]{{\it Institut f{\"u}r Theoretische
Physik, Physik-Department, Technische Universit{\"a}t M{\"u}nchen,
James-Franck-Stra{\ss}e, 85748 Garching bei M{\"u}nchen,
Germany}}\thanks{E-mail: {\tt tohlsson@physik.tu-muenchen.de}; Division
of Mathematical Physics, Theoretical Physics, Department of Physics,
Royal Institute of Technology (KTH), 100~44 Stockholm, Sweden},
{\large Walter
Winter}\addressmark[TUM]\thanks{E-mail: {\tt wwinter@physik.tu-muenchen.de}}}
     
\begin{abstract}
\noindent {\bf Abstract} \vspace{2.5mm}

In this paper, we show numerically that a symmetric Earth matter density
profile can, in principle, be reconstructed from a single baseline
energy spectrum up to a certain precision. For the numerical evaluations in
the high dimensional parameter space we use a genetic algorithm.

\vspace*{0.2cm}
\noindent {\it Key words:} neutrino oscillations, MSW effect, Earth's matter
density profile, neutrino tomography, genetic algorithm, geophysics\\
\noindent {\it PACS:} 14.60.Lm, 13.15.+g, 91.35.-x, 02.60.Pn
\end{abstract}

\maketitle

\section{Introduction}

In order to obtain knowledge about the Earth's interior, current measurements
are chiefly based on seismic wave propagation through the Earth (see, \eg,
\Refs~\cite{aki80,lay95}). The methods used in geophysics allow
a quite precise reconstruction of the seismic wave velocity profile. However,
for the determination of the matter density profile many assumptions have to
be made about the equation of state, relating the velocity and matter
density profiles. This process involves several uncertainties, since
additional information about the Earth's interior is quite rare (see, \eg,
\Refs \cite{Jean86,Jean90}). In neutrino physics, a method similar to X-ray
tomography has been suggested, namely
neutrino absorption tomography
\cite{Volkova74,DeRujula83,Wilson84,Askar84,Borisov87,Nicolaidis91,Crawford95,Kuo95,Jain:1999kp}. It requires many different baselines, since
neutrino absorption is not sensitive to the arrangement of the matter
structure along one baseline. Moreover, the cross section rises with
neutrino energy, which means that high energetic neutrinos are required.

Neutrino oscillations in the Earth have been extensively studied in general
and it has been suggested that neutrino oscillations in matter can
severely alter the energy spectrum of a neutrino beam through the Earth for
long baselines
\cite{Ermilova:1986ph,Baltz:1987hn,Nicolaidis:1988fe,Krastev:1988yu,Kuo:1989qe,Krastev:1989ix,Minorikawa:1990ip,Petcov:1998su,Akhmedov:1998ui,Akhmedov:1998xq,Chizhov:1999az,Chizhov:1999he,Akhmedov:1999ty,Mocioiu:2000st,Freund:2000ti,Dick:2000ce,Akhmedov:2000js}.
For quite realistic calculations of the transition probabilities the
Preliminary Reference Earth Model (PREM) or similar matter density profiles 
have been used \cite{stac77,dzie81}. In addition, several approximations of
the matter density profile, such as a step function (see, \eg,
\Refs~\cite{Freund:1999vc,Freund:1999gy}) or the first terms of a Fourier
series expansion \cite{Ota:2000hf}, turned out to supply rather good results. 
However, the inverse problem, \ie, what we can learn about the Earth's
structure from neutrino oscillations, is also considered to be an interesting
problem. In \Refs~\cite{Ermilova:1988pw,Chechin:1991} the approach of an
inverse scattering problem was suggested. However, transition probabilities of
a single baseline energy spectrum turned out not to be sufficient for
determining the matter density profile uniquely, \ie, additional information
on the relative phase of the output state vector was required. In this paper,
we ignore this information and investigate what we can learn about the Earth's
matter density profile from a single neutrino baseline energy spectrum using
two flavor neutrino oscillations in matter. We show that we can, in principle,
reconstruct the (symmetric) matter density profile, up to a certain precision
without any additional assumptions about the profile itself. The (large-scale)
symmetry of the matter density profile seems to be a reasonable assumption by
the results from geophysics. Furthermore, since two flavor neutrino
oscillations cannot distinguish time reverted matter density profiles (see,
\eg, \Refs~\cite{deGouvea:2000un,Akhmedov:2000cs}), omitting the symmetry
condition would {\it a priori} mean that reconstructed, asymmetric profiles
are not unique.

\section{The model}

For the numerical evaluation we use a (nondeterministic) genetic
algorithm \cite{Goldberg,Freeman}. It assumes an initial generation of random
matter density profiles and calculates their respective energy spectra.
The energy spectra are compared to a (realistic or measured) reference
spectrum by an error (fitness) function.
In this case, we choose minimization of a $\chi^2$-function \cite{Groom:2000in}
\begin{equation}
\chi^2 = \sum\limits_{i=1}^{B} 2 \left[ \langle x_i \rangle - x_i \right] +
2 x_i \log \frac{x_i}{\langle x_i \rangle},
\end{equation}
where $B$ is the
number of energy bins, $x_i$ the (integer) number of events in the $i$th bin,
and $\langle x_i \rangle$ the mean number of events in the $i$th bin.
Nondeterministic evolution of the matter density profiles over time
then creates an optimal or suboptimal matter density profile with an energy
spectrum similar to the reference spectrum. 
Since the whole process is
completely nondeterministic, the algorithm may find {\em any} symmetric matter
density profile which fits the energy spectrum with the required precision.
Thus, it also indicates if we are really able to obtain a unique solution
for the given energy spectrum. The strength of applying a nondeterministic
method is to reduce the exponential calculation effort to a polynomial one,
where the quality of the obtained result can be determined by the
statistical $\chi^2$-analysis. This means that we will find samples of the
(high dimensional) parameter space within a (high dimensional)
$n\sigma$-contour.

Here we use the neutrino appearance channel of a single
neutrino\footnote{Since for $\Delta m^2>0$ antineutrinos do not show resonant
behavior in matter, they are much less suitable for a reconstruction of the
matter density profile. Thus, we will restrict this discussion to neutrinos.}
baseline with an energy range from $2.5 \, \mathrm{GeV}$ to $20 \,
\mathrm{GeV}$ (an energy range which may be produced by a
neutrino factory), in order to cover the Mikheyev--Smirnov--Wolfenstein (MSW)
resonance regions for reasonable data \cite{mikh85,mikh86,wolf78}. Since we
are in principle interested in the reconstruction of the Earth's matter density
profile, we ignore information about the beam energy distribution and the cross
sections, \ie, we use the transition probabilities directly. However, it is
useful to have a statistical estimate of the significance of the results. Thus,
we assume a total number of events $I$ to be folded with the transition
probabilities equally spread over the whole energy spectrum at an equal cross
section. Furthermore, we use a two flavor neutrino oscillation scenario with a
symmetric PREM profile \cite{stac77,dzie81}. 
For the propagation of a neutrino state
vector along the baseline $L$ through the Earth, we divide the (symmetric)
matter density profile into $2N$ layers of equidistant length $\Delta L
\equiv \Delta L_k = L/(2N)$ of constant matter density $\rho_{k}$, where
$k=1,2,\hdots,2N$, and we use an evolution operator in flavor basis in each
individual layer $U_f(\Delta L_k,\rho_k)$ (see, \eg,
\Refs~\cite{Ohlsson:1999um,Ohlsson:2001et}).
The total evolution operator $U_f(L)$ is then the time-ordered product
of the ones in the individual layers, \ie, $U_f(L) = U_f(\Delta
L_{2N},\rho_{2N}) \ldots U_f(\Delta L_2,\rho_2) U_f(\Delta
L_1,\rho_1)$. Finally, the transition 
probabilities are given by the absolute values squared of the elements
of the total evolution operator, \eg, the transition probability for
$\nu_\alpha \to \nu_\beta$ is $P_{\alpha\beta}(L) = \vert \langle
\nu_\beta \vert U_f(L) \vert \nu_\alpha \rangle \vert$. Note that,
because of the symmetry, $N$ is the number of independent parameters
in this problem. The electron density $n_{e,k}$ is related to the
matter density $\rho_k$ by
\begin{equation}
n_{e,k} = \frac{Y_k \rho_k}{m_N}, \quad k=1,2,\hdots,2N,
\end{equation}
where $Y_k$ is the average number of electrons per nucleon in the
$k$th layer. In the Earth, $Y_k \simeq Y_{\rm Earth} \equiv 1/2$,
$k=1,2,\hdots,2N$, and $m_N$ is the 
nucleon mass.  For the baseline we choose $L=11736 \, \rm{km}$, corresponding
to a nadir angle of about $0.4$, in order to be able to see the Earth's mantle
as well as core in the energy spectra. Furthermore, for the oscillation
parameters we use $\Delta m^2 \equiv \Delta m_{32}^2 = 3.2 \cdot \, 10^{-3} \,
\mathrm{eV}^2$ \cite{Fukuda:2000np} and $\sin 2 \theta \equiv \sin 2
\theta_{13} = 0.1$ \cite{Apollonio:1998xe,Apollonio:1999ae}, which are
applicable for a mass hierarchy $\Delta m_{21}^2 \ll \Delta m_{32}^2$.

\section{Results and analysis}

We now present the results of some genetic algorithm
trial runs. For the number of steps $N$ in the matter density profile we
choose $N=4$, $N=9$, and $N=14$. We also compute the reference energy spectra
with the PREM profile divided into $2N$ layers of constant matter density for
$\chi^2 \equiv \Delta \chi^2$ to be equal to zero at the minimum. In almost
every example, we use a total number of events $I=100000$ to be folded with the
transition probabilities, corresponding to a very optimistic guess of a very
large neutrino factory, as well as a number of energy bins $B=31$. However, we
will also investigate the precision for a varying number of events and energy
bins.

\begin{figure*}[ht!]
\begin{center}
\includegraphics*[height=16cm]{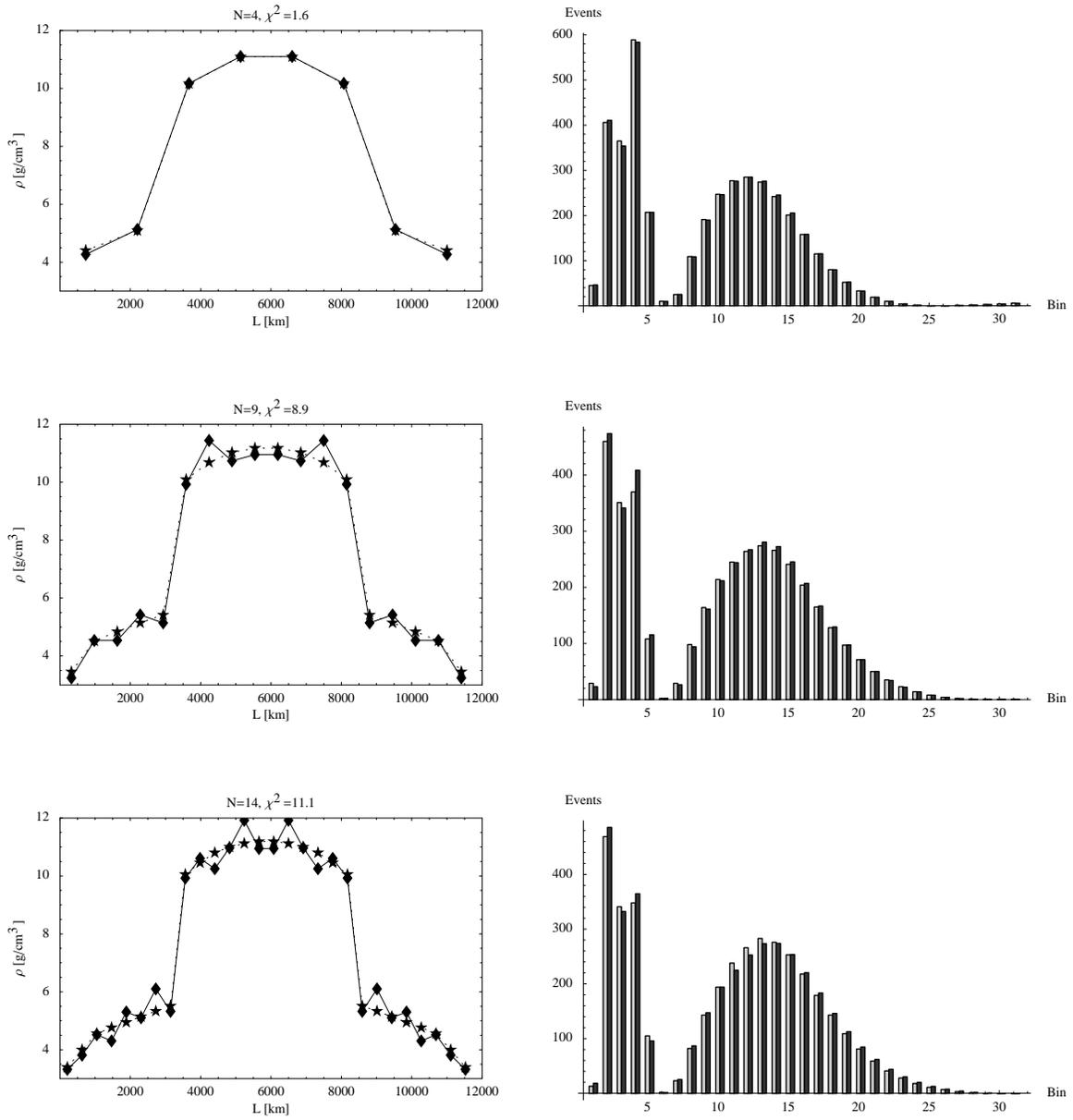}
\end{center}
\caption{\label{best} The best fits of the genetic algorithm trial runs for
$N=4$, $N=9$, and $N=14$ steps in the Earth's matter density profile. On the
left-hand side the best fit matter density profiles (solid lines and boxes)
are plotted with the reference profiles (dashed lines and stars). The values
of $N$ and $\chi^2$ are given above each plot. On the
right-hand side the energy bin spectra of the best fits (light bars) are shown
together with the spectra of the reference (dark bars).}
\end{figure*}

The best fit results of the genetic algorithm are shown in \fig~\ref{best}
for several numbers of steps $N$ of the matter density profile. Note that the
reference energy spectrum has two peaks corresponding to the mantle at high
energy and the core as well as interference effects at low energy,
respectively.  All results in this figure are within a $1 \sigma$ (68\%)
confidence region, which indicates that we cannot measure the matter density
profile preciser without additional assumptions. Therefore, the precision of
the measurement is limited for this type of experiment. Nevertheless, every
trial run of the algorithm converges to a matter density profile clearly
showing the mantle-core edge with high quality. It could not be seen in a
Fourier expansion approach, since many Fourier coefficients had to be
determined to obtain a high resolution of this edge \cite{Ota:2000hf}.

The fact that the initial profiles were completely created at random,
indicates that a single neutrino baseline energy spectrum really determines
the arrangement and matter densities of the large-scale
structure of the Earth's matter density profile.  Especially, in comparison to
X-ray or neutrino tomography, it can distinguish the order of the different
layers of matter because of the non-commuting operators used for the
propagation through the matter density profile. The algorithm may have found
any matter density profile matching the reference profile, such as one with
mantle and core exchanged. Nevertheless, it turned out that it converged
in all cases against the well-known mantle-core-mantle structure of the Earth.
However, the non-commuting operators make the analytical inversion of the
energy spectrum quite complicated (\cf,
\Refs~\cite{Ermilova:1988pw,Chechin:1991}). 

\begin{figure*}[ht!]
\begin{center}
\includegraphics*[height=11.5cm]{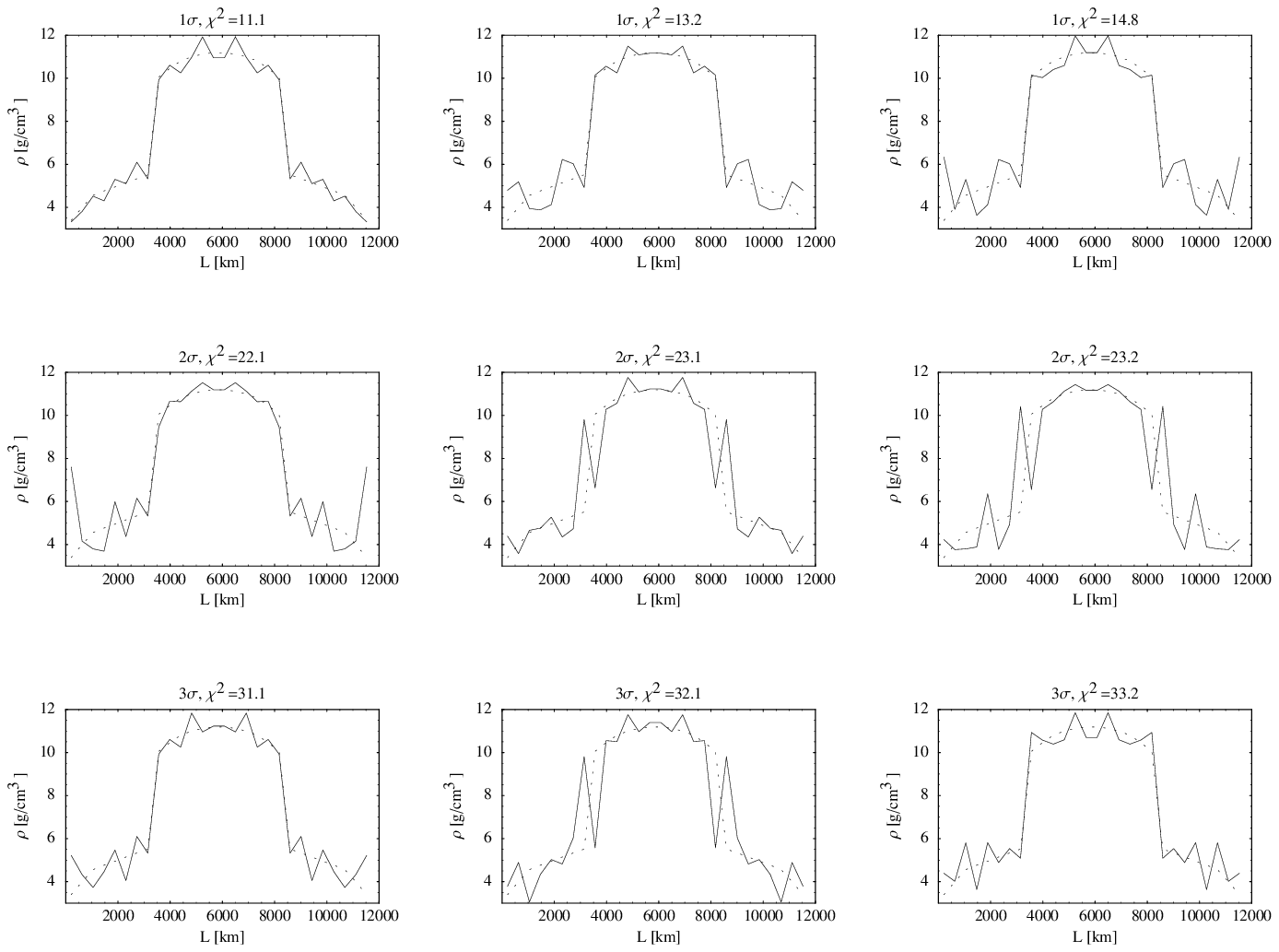}
\end{center}
\caption{\label{samplespace} Representatives of the $N=14$-dimensional
parameter space for $1 \sigma$ (first row: $\chi^2 \le 15.9$), $2 \sigma$
(second row: $\chi^2 \le 24.0$), and $3 \sigma$ (third row: $\chi^2 \le 33.2$)
confidence levels. The respective $\chi^2$'s of the samples are given
above each plot. The sample matter density profiles are
plotted with solid lines, whereas the reference profile is plotted with dashed
lines.} \end{figure*}

As already mentioned above, the resolution of the matter density profile
seems to be bounded. In order to investigate this, let us take a
closer look at the parameter space. Since it is impossible to visualize the
contours of equal confidence levels of a high dimensional problem,
we plotted in \fig~\ref{samplespace} for $N=14$ some representatives of the
parameter space
within the $1 \sigma$, $2
\sigma$, and $3 \sigma$ contours. Again, the
mantle-core edge can be easily resolved at the $1 \sigma$-level, but in many
cases not for lower precision. It
turns out that this effect grows with an increasing number of steps in the
matter density profile. In addition, \figs~\ref{best} and \ref{samplespace}
indicate that the matter density profile can be quite well reconstructed for a
small number of steps $N$, but for a higher number of steps
$N$ small fluctuations enter the reconstructed profile. As can be seen in
\fig~\ref{samplespace}, we especially observe for $N=14$ small fluctuations in
the mantle and core regions, indicating bounds on the spatial resolution. In
\App~\ref{sec:res}, we show analytically with a perturbation theoretical
method what is intuitively clear: structures of small amplitude on a length
scale $\lambda$ much smaller than the oscillation length in matter
$L^{\mathrm{osc}}_{\mathrm{matter}} = 4 \pi E/\Delta \tilde{m}^2 = 4 \pi
E/[\xi(n_e) \Delta m^2]$ cannot be resolved by neutrino oscillation in matter.
Here \begin{eqnarray}
\xi(n_{e}) &\equiv &\sqrt{\left( \frac{2 \sqrt{2} E G_{F} n_{e}}{\Delta
m^2} - \cos 2 \theta \right)^2 + \sin^2 2 \theta} \nonumber \\
 \label{xi}
\end{eqnarray}
is determined by the mean matter density $n_e$.
Since $\xi \ge | \sin 2 \theta |$, the upper bound
$\lambda \simeq L^{\mathrm{osc}}_{\mathrm{matter}} < 4 \pi E/(\sin 2 \theta
\Delta m^2) \propto E$ indicates that the higher the resonance energy of the
resonance peak is, the lower the spatial resolution becomes. For this reason,
the resolution of the core, corresponding to the low energy peak, is probably
higher (\cf, \figs~\ref{best} and \ref{samplespace}). However, for a realistic
neutrino factory one has suppressions in low energetic events, which means that
this effect is probably compensated by statistics \cite{Barger:1999fs}.

\begin{figure*}[ht!]
\begin{center}
\includegraphics*[bb = 0 0 360 750,height=16cm,angle=270]{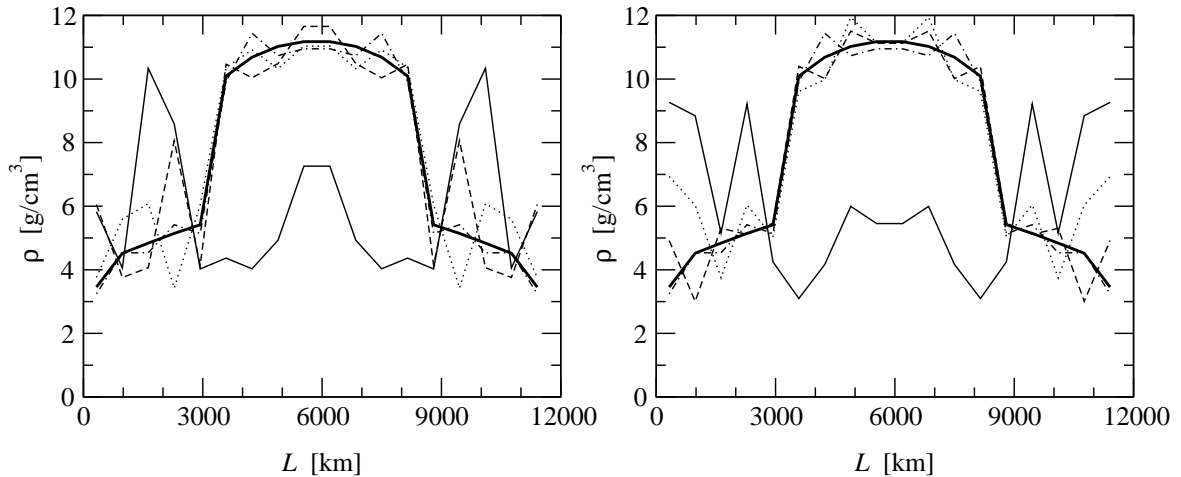}
\end{center}
\vspace*{-1cm}
\caption{\label{eventbindep} Matter density profiles for different numbers of
total events $I$ (left) and different numbers of energy bins $B$ (right). The
figures show matter density profiles close to the $1 \sigma$-contour ($\chi^2
\simeq 10.42$) for $N=9$ steps.
The following matter density curves are plotted: In the left plot ($B=31$):
Reference (thick solid); $I=1000$ (thin solid); $I=5000$ (dotted); $I=10000$
(dashed); $I=100000$ (dashed-dotted). In the right plot ($I=100000$): Reference
(thick solid); $B=5$ (thin solid); $B=9$ (dotted); $B=16$ (dashed); $B=31$
(dashed-dotted).} \end{figure*}

In order to investigate the dependence on the total number of events $I$ and
energy bins $B$, we show in \fig~\ref{eventbindep} some matter density profile
examples close to the $1 \sigma$-contour for different numbers of $I$ and $B$.
 One can see that the precision increases with
the number of events. In almost all cases, we can clearly resolve the
mantle-core edge, but for too few events $I \ll 5000$ the result does not have
a physical meaning anymore.  One can also see that the precision increases
with the number of energy bins $B$. In addition, for $B=5$ the matter density
profile again looses its physical meaning. Note that, since we did not
integrate over the energy of each bin, but used the mean energy, there is a
statistical error increasing with decreasing number of energy bins. However,
one may expect a natural boundary of the evaluation for $B < N$, because $N$
represents the number of unknowns and $B$ the number of equations. This
equivalence between the number of equations and the number of energy bins
would be destroyed for an integration over the energy within each energy bin.

\section{Summary and conclusions}

In summary, we have demonstrated that we can, in principle,
reconstruct the symmetric Earth matter density profile using a single
neutrino baseline energy spectrum without additional assumptions. However, it
turned out that the precision is limited in this method, especially since we
made quite optimistic assumptions about the total number of detectable events
$I$ and the energy spectrum of the source, as well as we ignored any energy
smearing.

We conclude that
neutrino physics cannot be used as the only source for obtaining information
about the Earth's interior, since the precision is presently lower
than in geophysics, and the resolution has a natural lower bound, the
oscillation length in matter. However, geophysics can only access the velocity
profile of seismic waves directly. Additional assumptions about the equation
of state have to be made in order to access the matter density profile. Thus,
neutrino oscillations in matter could help to test and verify the equation of
state. Similarly, the parameter space in neutrino reconstruction tomography
could be shrinked by using knowledge from geophysics.

\section*{Acknowledgements}

We would like to thank Evgeny Akhmedov, Martin Freund, Patrick Huber, and
Manfred Lindner for useful discussions and comments.

This work was supported by the Swedish Foundation for International
Cooperation in Research and Higher Education (STINT) [T.O.], the Wenner-Gren
Foundations [T.O.], the ``Studienstiftung des deutschen Volkes'' (German
National Merit Foundation) [W.W.], and the ``Sonderforschungsbereich 375
f{\"u}r Astro-Teilchenphysik der Deutschen Forschungsgemeinschaft''.

\begin{appendix}

\section{Spatial resolution of neutrino oscillations in matter}
\label{sec:res}

In this appendix, we will show in the case of two neutrino flavors that we
cannot resolve structures much smaller than the oscillation length in matter.
For that we state the differential equation describing the time evolution of
flavor states in matter:
\begin{equation}
 i \frac{d}{dt} \nu_f(t) = \mathcal{H}_0 \, \nu_f(t) + 
\mathcal{H}_{\mathrm{int}}(n_e) \, \nu_f(t)
 \label{tevol}
\end{equation}
with $\nu_f(t) = (\nu_e(t), \nu_{\mu}(t) )^T$,
\begin{equation}
\mathcal{H}_0 = \left(
\begin{array}{cc}
 -\frac{\Delta m^2}{4E} \cos 2 \theta & \frac{\Delta m^2}{4E} \sin 2 \theta \\
 \frac{\Delta m^2}{4E} \sin 2 \theta & \frac{\Delta m^2}{4E} \cos 2 \theta \\
\end{array} \right),
\end{equation}
and
\begin{equation}
\mathcal{H}_{\mathrm{int}}(n_e) = \left(
 \begin{array}{cc}
 \sqrt{2} G_F n_e & 0 \\ 0 & 0
\end{array} \right).
\end{equation}
Let us introduce a perturbation $\eta(t)$ on a shorter timescale than a
slowly varying matter density profile $\tilde{n}_e(t)$, \ie,
\begin{equation}
 n_e(t) = \tilde{n}_e(t)+\eta(t)
 \label{pertne}
\end{equation}
with $\eta(t)$ making many oscillations while $\tilde{n}_e(t)$ is
approximately constant. This will also result in a modification of the
amplitudes
\begin{equation}
 \nu_f(t) = \tilde{\nu}_f(t) + \chi_f(t),
 \label{pertnu}
\end{equation}
where $\tilde{\nu}_f(t)$ solves the unperturbed equations for
$\mathcal{H}_{\mathrm{int}}(\tilde{n}_e)$.
Thus, we can split \eq~(\ref{tevol}) by applying \eqs~(\ref{pertne})
and (\ref{pertnu}) and we obtain
\begin{eqnarray}
i \frac{d}{dt} \tilde{\nu}_f(t) & = & \mathcal{H}_0 \, \tilde{\nu}_f(t) + 
\mathcal{H}_{\mathrm{int}}(\tilde{n}_e(t)) \, \tilde{\nu}_f(t),\nonumber\\ \\
 i \frac{d}{dt} \chi_f(t) & = & \mathcal{H}_0 \, \chi_f(t)
+\mathcal{H}_{\mathrm{int}}(\tilde{n}_e(t)) \, \chi_f(t) \nonumber \\
&+& \mathcal{H}_{\mathrm{int}}(\eta(t)) \, \tilde{\nu}_f(t).
 \label{osc2}
\end{eqnarray}
Here we have neglected second order corrections of the order $\eta(t) \chi(t)$.
We know that the oscillation of the solution of the first equation is
determined by the oscillation length in matter
$L^{\mathrm{osc}}_{\mathrm{matter}} = 4 \pi E/\Delta \tilde{m}^2 = 4 \pi
E/[\xi(\tilde{n}_e) \Delta m^2]$, where the $\xi$ was defined in
\eq~(\ref{xi}) and refers here to the constant mean matter density
$\tilde{n}_e$.\footnote{For the approximation of a matter density profile
by its mean, see \Ref \cite{Ohlsson:2001et}.}. For the second equation, we may
assume a quick periodic oscillation $\eta(t) = A \cos( \omega t)$ with the
amplitude $A$ and the frequency $\omega$ on a timescale much shorter than the
one of the first equation. Hence, we can in the second equation take
$\tilde{\nu}_f(t) \simeq \tilde{\nu}_f$ and $\tilde{n}_e(t) \simeq \tilde{n}$
approximately to be constant. 

We can thus
rewrite \eq~(\ref{osc2}) as
\begin{eqnarray} 
i \frac{d}{dt} \chi_f(t) & = & \mathcal{H}_0 \, \chi_f(t)
+\mathcal{H}_{\mathrm{int}}(\tilde{n}_e) \, \chi_f(t) \nonumber \\
&+& \mathcal{H}_{\mathrm{int}}(\tilde{\nu}_e) \, A \cos(\omega t).
\end{eqnarray}
This is a differential system of equations
similar to the one for two coupled oscillators, where the one representing
the electron flavor is driven by an external force corresponding to the
oscillatory fluctuation in the matter density profile. It is solved by quite
lengthy, purely oscillatory expressions for $\chi_e(t)$ and $\chi_{\mu}(t)$.

One can show that the
amplitudes of the oscillations reduce to zero when $\omega \gg
\xi(\tilde{n}_e) \, \Delta m^2/(2 E) \equiv 2 \pi /
L^{\mathrm{osc}}_{\mathrm{matter}}$. Since $\omega = 2 \pi/\lambda$, this
means that we cannot resolve structures in the matter density profile with a
width $\lambda \ll L^{\mathrm{osc}}_{\mathrm{matter}}$. Note that the
perturbation method used breaks down for too large amplitudes or too slow
oscillations.

\end{appendix}

\bibliographystyle{h-elsevier}
\bibliography{references}

\end{document}